\RequirePackage{fix-cm}
\documentclass{svjour3}                     % onecolumn (standard format)
\smartqed  % flush right qed marks, e.g. at end of proof
\usepackage{graphicx}
\usepackage{mathptmx}      % use Times fonts if available on your TeX system
\begin{document}

\title{
Application of Time Transfer Function to
McVittie Spacetime:\\
Gravitational Time Delay and
Secular Increase in Astronomical Unit
%\thanks{Grants or other notes
%about the article that should go on the front page should be
%placed here. General acknowledgments should be placed at the end of the article.}
}
%\subtitle{}

%\titlerunning{Short form of title}        % if too long for running head

\author{Hideyoshi Arakida
}

%\authorrunning{Short form of author list} % if too long for running head

\institute{H. Arakida \at
School of Education, Waseda University\\
1-6-1, Nishiwaseda, Shinjuku 169-8050 Tokyo JAPAN\\
\email{arakida@edu.waseda.ac.jp}           %  \\
%%             \emph{Present address:} of F. Author  %  if needed
}

%\date{Received: date / Accepted: date}
% The correct dates will be entered by the editor

\maketitle

\begin{abstract}
We attempt to calculate the gravitational time delay 
in a time-dependent gravitational field, especially in 
McVittie spacetime, which can be considered as the 
spacetime around a gravitating body such as the Sun, embedded in 
the FLRW (Friedmann-Lema\^itre-Robertson-Walker) cosmological 
background metric. To this end, we adopt the time transfer 
function method proposed by Le Poncin-Lafitte {\it et al.} 
(Class. Quant. Grav. 21:4463, 2004) and Teyssandier and 
Le Poncin-Lafitte (Class. Quant. Grav. 25:145020, 2008), 
which is originally related to Synge's world function 
$\Omega(x_A, x_B)$ and enables to circumvent the integration of
the null geodesic equation. We re-examine the  global cosmological effect 
on light propagation in the solar system.  
The round-trip time of a light ray/signal is given by the 
functions of not only the spacial coordinates but also the 
emission time or reception time of light ray/signal, which 
characterize the time-dependency of solutions. We also apply the
obtained results to the secular increase in the astronomical unit,
reported by Krasinsky and Brumberg 
(Celest. Mech. Dyn. Astron. 90:267, 2004), and we show that the 
leading order terms of the time-dependent component due to 
cosmological expansion is 9 orders of magnitude smaller 
than the observed value of $d{\rm AU}/dt$, 
i.e., $15 \pm 4$ ~[m/century].
Therefore, it is not possible to explain the secular increase in 
the astronomical unit in terms of cosmological expansion.
%%%%%%%%%%%%%%%%%%%%%%%%%%%%%%%%%%%%%%%%%%%%%%%%%%%%%%%%%%%%%%
% \PACS{PACS code1 \and PACS code2 \and more}
% \subclass{MSC code1 \and MSC code2 \and more}
\end{abstract}
%%%%%%%%%%%%%%%%%%%%%%%%%%%%%%%%%%%%%%%%%%%%%%%%%%%%%%%%%%%%%%%%%%%%%
\section{Introduction\label{intro}}
%%%%%%%%%%%%%%%%%%%%%%%%%%%%%%%%%%%%%%%%%%%%%%%%%%%%%%%%%%%%%%%%%%%%%
Technological progress has significantly enhanced the 
accuracy of solar system observations. Presently, 
to archive an accuracy of 9 to 11 digits or higher, the
1st order ${\cal O}(c^{-2})$ or ${\cal O}(G)$ 
($c$ is the speed of light in vacuum and $G$ is the gravitational 
constant) post-Newtonian/post-Minkowskian effects of general relativity 
must be incorporated in a number of observational models. 
Moreover, several space missions such as GAIA \cite{gaia}, 
Laser Astrometric Test Of Relativity (LATOR) \cite{lator}, and 
Astrodynamical Space Test of Relativity using Optical Devices (ASTROD) 
\cite{astrod} are planed and especially LATOR and ASTROD can potentially 
detect post-linear or 2nd order ${\cal O}(c^{-4})$ or ${\cal O}(G^2)$ 
post-Newtonian/post-Minkowskian effects.

Highly accurate and rigorous light propagation models 
are required to analyze high precision data such as
the radar time delay, deflection of light rays, and
frequency shift of radio signals. 
Although, these effects have been extensively investigated
(see i.e.
\cite{kopeikin1999,kopeikin2002,klioner2003a,klioner2003b,defellice2004,defelice2006}
and references therein, 
most of the previous studies were based on integrating the null 
geodesic equation of light rays. This approach is effective in the
case of linear or 1st order approximation. However, it is generally 
difficult to compute the contribution of 2nd and
higher order effects, even in the case of static gravitational 
fields; the situation becomes further complicated in the case of 
time-dependent spacetime.

A novel method known as the time transfer function method 
has been recently developed 
\cite{plt2004,pt2008} to overcome such circumstances.
This approach is originally based on Synge's world function
$\Omega(x_A, x_B)$ \cite{synge},
\begin{equation}
 \Omega(x_A, x_B) \equiv 
  \frac{1}{2}(\lambda_B - \lambda_A)
  \int^{\lambda_B}_{\lambda_A}
  g_{\mu\nu}
  \frac{dx^{\mu}}{d\lambda}
  \frac{dx^{\nu}}{d\lambda}
  d\lambda,
  \label{wf1}
\end{equation}
where $g_{\mu\nu}$ is a metric tensor of spacetime,
$x_A = (x^0_A = ct_A, x^i_A = \vec{x}_A)$ and 
$x_B = (x^0_B = ct_B, x^i_B = \vec{x}_B)$ are the coordinates 
of end points $A$ and $B$, respectively, on the geodesic world-line,
and $\lambda$ is the affine parameter. 
Then, the world function $\Omega(x_A, x_B)$ is 
defined as the half length of the world line between $A$ and $B$.
On the basis of (\ref{wf1}), the time transfer functions 
introduced in \cite{plt2004,pt2008} define
the travel time of a light rays/signal between two points,
i.e., $t_B - t_A$.
\cite{plt2004} and \cite{pt2008} succeeded in constructing the general 
$N$-th order post-Minkowskian 
approximation scheme with respect to the gravitational constant $G$
via an iterative procedure, and they showed that
this method can reduce the number of calculations by 
taking the static Schwarzschild spacetime for instance.

On the other hand, in the field of experimental relativity, 
astrometric observations in the solar system have played 
a crucial role in the verification of gravity. It has been 
show that the main PPN (parametrized post-Newtonian) parameters 
$\beta$ and $\gamma$ are equal to unity for general relativity, 
($\beta = \gamma = 1$); as increasing accuracy,
$\beta$ and $\gamma$ are strictly constrained to unity
(see Fig. 1 of \cite{turyshev2009}).
The enhanced accuracy of observations accounts for the existence 
of unexplained phenomena by means of the framework of general
relativity; such phenomena include the anomalous acceleration 
of the Pioneer spacecraft \cite{tt2010}, Earth flyby anomaly 
\cite{anderson2008}, 
the anomalous perihelion precession of Saturn \cite{iorio2009}, 
and the secular increase in the astronomical unit \cite{kb2004}. 
Presently, the origins of these anomalies are far from clear;
however, they may be attributable to some fundamental properties 
of gravitation (see \cite{anderson2010,lpd2008} and 
the references therein).

The secular increase in the astronomical unit is particularly
interesting. Here, the astronomical unit implies that the conversion 
constant between two length units and the current best-fit value is 
\cite{pitjeva2005}
\begin{eqnarray}
 \frac{1 ~[{\rm AU}]}{1 ~[{\rm m}]} \equiv
  {\rm AU} =
   1.495978706960 \pm 0.1.
   \label{au1}
\end{eqnarray}
The determination of astronomical unit is directly related to 
the light propagation formula, i.e., the equation of light time. 
Without loss of generality, the secular increase in the astronomical
unit was found by the relation
\begin{eqnarray}
 t_{\rm theo} = \frac{d_{\rm theo}}{c}
  \left[
   {\rm AU} + \frac{d{\rm AU}}{dt}(t - t_0)
  \right] ~~~~[{\rm s}],
  \label{au2}
\end{eqnarray}
where $t_{\rm theo}$ is the computed round-trip time (theoretical value), 
$d_{\rm theo}$ is the interplanetary distance obtained 
from planetary ephemerides (note that the length unit of $d_{\rm theo}$ 
is [AU]), ${\rm AU}$ and $d{\rm AU}/dt$ are the astronomical unit and 
its time variation, respectively, and $t_0$ is the initial epoch.
$t_{\rm theo}$ is compared with the observed round-trip time 
$t_{\rm obs}$. Krasinsky and Brumberg estimated 
$d{\rm AU}/dt$ as $15 \pm 4$ [m/cy]
\footnote{In this paper, cy denotes century 
according to Krasinsky and Brumberg (2004).}.

The influence of cosmological expansion has been investigated
as a potential cause of secular variation in AU
\cite{kb2004,mashhoon,arakida2009}. In particular, 
\cite{kb2004} and \cite{arakida2009} examined
its contribution to light propagation. The approach of \cite{kb2004} 
is somewhat qualitative, and the discussion of \cite{arakida2009} is 
essentially reduced to a static model that is equivalent 
to Schwarzschild-de Sitter spacetime.
These circumstances are due to the fact that 
the time-dependent null geodesic equation must be solved
in order to study the cosmological effect on light propagation;
however it is generally difficult to solve this equation.
Nonetheless, the time transfer function method proposed in
\cite{plt2004,pt2008} is applicable not only to the static case 
but also to the time-dependent case. Therefore, it is useful to 
re-examine the effect of cosmological expansion on the basis of the 
time transfer function and to compare the theoretically obtained 
results with the observed $d{\rm AU}/dt$.

In this paper, according to the time transfer function method,
we consider light propagation in a time-dependent gravitational 
field, and we focus our discussion on the gravitational time delay. 
We adopt the McVittie model \cite{robertson,mcvittie} 
as the time-dependent spacetime; it can be considered as the spacetime 
around a gravitating body such as the Sun, embedded in the FLRW 
(Friedmann-Lema\^itre-Robertson-Walker) cosmological background metric. 
The time-dependency in the solution is characterized by 
either the emission time or the reception time of a light ray/signal. 
Then, the obtained results is applied to the secular (time) variation 
in the astronomical unit, reported in \cite{kb2004}.

The remainder of this paper is organized as follows. 
In Section \ref{ttfm}, we briefly summarize the time transfer 
function, and in Section \ref{rms}, we investigate the 
gravitational time delay in McVittie spacetime. 
In Section \ref{siau}, we discuss the secular increase in the 
astronomical unit. Finally, in Section \ref{concl}, we 
conclude the paper with a summary of our study.
%%%%%%%%%%%%%%%%%%%%%%%%%%%%%%%%%%%%%%%%%%%%%%%%%%%%%%%%%%%%%%%%%%%%%%%%%
\section{Time Transfer Function\label{ttfm}}
%%%%%%%%%%%%%%%%%%%%%%%%%%%%%%%%%%%%%%%%%%%%%%%%%%%%%%%%%%%%%%%%%%%%%%%%%
Before discussing the gravitational time delay in McVittie spacetime, 
we shall briefly summarize the time transfer function method 
(see \cite{plt2004,pt2008} for the details).

It is generally difficult to acquire the world function 
$\Omega(x_A, x_B)$ concretely; nevertheless, in the case of 
Minkowskian flat spacetime, the world function is easily obtained by 
using the parameter equation $x(\lambda) = (x_B - x_A)\lambda + x_A$ and 
by setting $\lambda_A = 0$ and $\lambda_B = 1$
\cite{plt2004,synge}:
\begin{equation}
 \Omega^{(0)}(x_A, x_B) =
\frac{1}{2}\eta_{\mu\nu}
(x^{\mu}_B - x^{\mu}_A)
(x^{\nu}_B - x^{\nu}_A),
\label{wf2}
\end{equation}
where the $x^{\mu}$ ($\mu = 0, 1, 2, 3$) are Minkowskian
coordinates with respect to the Minkowski metric
$\eta_{\mu\nu}= {\rm diag}(-1; 1; 1; 1)$.
For the null geodesic, the world function $\Omega(x_A, x_B)$ 
satisfies the condition
\begin{equation}
 \Omega(x_A, x_B) = 0
  \label{null1}
\end{equation}
because $ds^2 = 0$. Hence, from (\ref{wf2}) and (\ref{null1}),
the travel time between $A$ and $B$, i.e., $t_B - t_A$,
in Minkowski spacetime becomes
\begin{equation}
 c^2(t_B - t_A)^2 = \delta_{ij}
  (x^i_B - x^i_A)
  (x^j_B - x^j_A) = R_{AB}^2,
  \label{wf3}
\end{equation}
where $\delta_{ij}$ is Kronecker's delta.
The time transfer function approach starts from (\ref{wf3}), 
and the weak-field approximation is recursively developed
with respect to the gravitational constant $G$.

If the metric has the form
\begin{eqnarray}
 g_{\mu\nu} = \eta_{\mu\nu} + h_{\mu\nu},\quad
  |h_{\mu\nu}| \ll 1,
\end{eqnarray}
where $h_{\mu\nu}$ is a perturbation to $\eta_{\mu\nu}$, 
the time transfer functions that give the travel time of
the light ray/signal are formally expressed as follows:
\begin{eqnarray}
 t_B - t_A &=&
  {\cal T}_e(t_A, \vec{x}_A, \vec{x}_B) 
  =
  \frac{1}{c}
  \left[
  R_{AB} + \Delta_e(t_A, \vec{x}_A, \vec{x}_B)
  \right]
  \label{ttf1}\\
  &=& 
  {\cal T}_r(\vec{x}_A, t_B, \vec{x}_B)
  =
  \frac{1}{c}
  \left[
  R_{AB} + \Delta_r(\vec{x}_A, t_B, \vec{x}_B)
  \right],
  \label{ttf2}
\end{eqnarray}
where ${\cal T}_e(t_A, \vec{x}_A, \vec{x}_B)$ is the
emission time transfer function in spacial coordinates 
$\vec{x}_A, \vec{x}_B$ and $t_A$ is the emission time of the signal,
${\cal T}_r(\vec{x}_A, t_B, \vec{x}_B)$ is the reception time 
transfer function in spacial coordinates $\vec{x}_A, \vec{x}_B$ and
$t_B$ is the reception time of the signal, 
$R_{AB} = |\vec{x}_B - \vec{x}_A|$, and $\Delta_e$ and $\Delta_r$ are 
the emission time delay function and reception time delay function, 
respectively. $\Delta_e$ and $\Delta_r$ characterize the 
gravitational time delay. In (\ref{ttf1}) and (\ref{ttf2}), 
$R_{AB}$ comes from (\ref{wf3}). 
Henceforth, $A$ denotes the emission and $B$ denotes the reception.

$\Delta_e$ and $\Delta_r$ can be iteratively calculated up to
any order of approximation \cite{plt2004,pt2008}; 
nevertheless, in this paper, we only need the 1st order formulae,
i.e.,
\begin{eqnarray}
	\Delta^{(1)}_e &=&
	 \frac{R_{AB}}{2}\int^1_0
	 \left[
	  g^{00}_{(1)} - 2N^i_{AB}g^{0i}_{(1)} + 
	  N^i_{AB}N^j_{AB}g^{ij}_{(1)}
	 \right]d\mu
	 \label{tdf1}\\
	\Delta^{(1)}_r &=&
	 \frac{R_{AB}}{2}\int^1_0
	 \left[
	  g^{00}_{(1)} - 2N^i_{AB}g^{0i}_{(1)} + 
	  N^i_{AB}N^j_{AB}g^{ij}_{(1)}
	 \right]d\lambda,
	 \label{tdf2}
\end{eqnarray}
where $g^{\mu\nu}_{(1)}$ indicates the 1st order perturbation
with respect to $\eta^{\mu\nu}$ and $N^i_{AB} = (x^i_B - x^i_A)/R_{AB}$.
In (\ref{tdf1}) and (\ref{tdf2}), integration is 
carried out along the straight line
\begin{eqnarray}
	& &t(\mu) = t_A + \mu T_{AB}, \quad
	x(\mu) = x_A + \mu (x_B - x_A)
	~~~~\mbox{for}~~~~\Delta^{(1)}_e,
	\label{pe1}\\
 	& &t(\lambda) = t_B - \lambda T_{AB}, \quad
	x(\lambda) = x_B - \lambda (x_B - x_A)
	~~~~\mbox{for}~~~~\Delta^{(1)}_r,
	\label{pe2}
\end{eqnarray}
where $T_{AB}$ is the time lapse between $A$ and $B$ along the
straight line. Then, we can put $T_{AB} = R_{AB}/c$. 
%%%%%%%%%%%%%%%%%%%%%%%%%%%%%%%%%%%%%%%%%%%%%%%%%%%%%%%%%%%%%%%%%%%%%%
\section{Gravitational Time Delay in McVittie Spacetime
\label{rms}}
%%%%%%%%%%%%%%%%%%%%%%%%%%%%%%%%%%%%%%%%%%%%%%%%%%%%%%%%%%%%%%%%%%%%%%
\subsection{McVittie Spacetime}
%%%%%%%%%%%%%%%%%%%%%%%%%%%%%%%%%%%%%%%%%%%%%%%%%%%%%%%%%%%%%%%%%%%%%%
The McVittie metric is expressed in standard comoving 
form \cite{robertson,mcvittie} as
\begin{eqnarray}
 ds^2 = -
  \left[
   \frac{1 - \frac{GM}{2c^2ra(t)}}{1 + \frac{GM}{2c^2ra(t)}}
  \right]^2 c^2dt^2
  + \left[1 + \frac{GM}{2c^2ra(t)}\right]^4
  a^2(t)(dr^2 + r^2d \Omega^2),
  \label{rm1}
\end{eqnarray}
where $d\Omega^2 = d\theta^2 + \sin^2 \theta d\phi^2$,
$M$ is the mass of the central gravitating body, and $a(t)$ is a 
scale factor. (\ref{rm1}) reduces to the Schwarzschild solution when
$a(t) = {\rm constant}$, and it reduces to the FLRW cosmological 
model for the curvature parameter $k = 0$ when $M = 0$.

The various observational models of the solar system are 
currently formulated in some kind of proper coordinate system
such as the barycentric celestial reference system (BCRS) based on 
the post-Newtonian framework \cite{soffel2003}, instead of 
the cosmological comoving frame. Hence, to compare the effects 
formulated using the proper coordinates with the cosmological 
ones within the same framework, we adopt the radial 
transformation
\cite{robertson,jarnefelt1,jarnefelt2,nolan1,nolan2,carrera}
\begin{eqnarray}
 R = a(t)r
  \left[
   1 + \frac{GM}{2c^2ra(t)}
  \right]^2,
\end{eqnarray}
then (\ref{rm1}) is rewritten as
\begin{eqnarray}
 ds &=& - \left(1 - \frac{2GM}{c^2 R}\right)
  c^2 dt^2
  + \left(
     \frac{dR}{\sqrt{1 - \frac{2GM}{c^2 R}}} - \frac{HR}{c}cdt
    \right)^2 + R^2d\Omega^2
  \nonumber\\
  &=& - \left(1 - \frac{2GM}{c^2 R} - \frac{H^2 R^2}{c^2}\right)
  c^2 dt^2 - \frac{2HR}{c}\left(1 + \frac{GM}{c^2 R} + 
			  {\cal O}(M^2)\right)cdt dR
  \nonumber\\
& &+ \left(1 + \frac{2GM}{c^2 R} + {\cal O}(M^2)\right)dR^2 + R^2d\Omega^2,
\label{rm2}
\end{eqnarray}
where $H = H(t) = \dot{a}(t)/a(t)$ is the Hubble parameter.

\begin{figure}
   \begin{center}
    \includegraphics[scale=0.4]{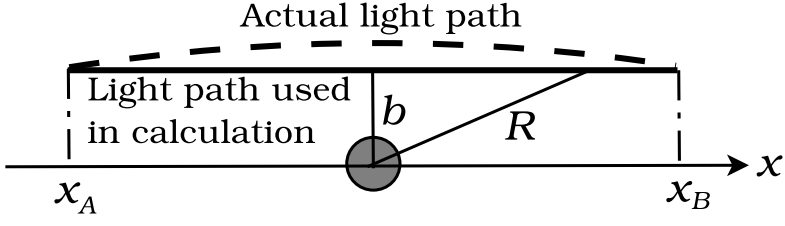}
   \end{center}
 \caption{Light ray/signal path. We assume that the path 
 used in calculation is rectilinear and parallel to the $x$-axis 
 (bold solid line); $b$ is the impact parameter
 and $R = \sqrt{x^2 + b^2}$. The actual path is denoted by the 
 bold dashed line. \label{fig1}}
  \end{figure}

As mentioned in previous section, the light path used in the
computation is rectilinear so that the rectangular coordinate system 
can be used instead of  the spherical coordinate system.
By coordinate transformation,
\begin{eqnarray}
 x = R\sin\theta\cos\phi,\quad
  y = R\sin\theta\sin\phi,\quad
  z = R\cos\theta,
\end{eqnarray}
and (\ref{rm2}) becomes (see \cite{brumberg2})
 \begin{eqnarray}
  ds^2 &=& 
   - \left(1 - \frac{2GM}{c^2 R} - \frac{H^2 R^2}{c^2}\right)c^2 dt^2
- \frac{2Hx^i}{c}\left(1 + \frac{2GM}{c^2 R} +
		 {\cal O}(M^2)\right)cdt dx^i
\nonumber\\
& &+ \left(\delta_{ij} + \frac{2GM}{c^2 R^3}x^i x^j + {\cal O}(M^2)
     \right)dx^i dx^j,
   \label{rm3}
 \end{eqnarray}
where $R = \sqrt{x^2 + y^2 + z^2}$. 
To simplify the computation, 
the straight line used in integration is parallel to the $x$-axis 
(see Fig. \ref{fig1}). Hence, (\ref{rm3}) reduces to
 \begin{eqnarray}
  ds^2 &=& - \left(1 - \frac{2GM}{c^2 R} - \frac{H^2 R^2}{c^2}\right)
   c^2 dt^2 - \frac{2Hx}{c}
    \left(1 + \frac{2GM}{c^2 R} + {\cal O}(M^2)\right)cdt dx
    \nonumber\\
  & &
   + \left(1 + \frac{2GM}{c^2 R^3} x^2 + {\cal O}(M^2)
     \right)dx^2,
   \label{rm4}
 \end{eqnarray}
where $y = b = {\rm constant}$ ($b$ is the impact factor),
$z = 0$, $R = \sqrt{x^2 + b^2}$, and in this case, we may put
$N^x_{AB} = (x_B - x_A)/R_{AB}, N^y_{AB} = N^z_{AB} = 0$.
This approach is similar to the one described in \cite{mtw}, 
(see Section 40.4 and Fig. 40.3 of \cite{mtw}).
In (\ref{rm4}), we see the Hubble parameter $H(t)$, 
which generally is an arbitrary function of time $t$. 
Hence, we suppose that $H(t)$ changes adiabatically because the time
lapse of a light ray/signal observed in the solar system is much shorter 
than the age of the universe, $T_{\rm U} \approx 10^{10}~[{\rm yr}]$: 
\begin{equation}
 H(t) = H_0 + \left.\frac{dH}{dt}\right|_0 t
  + {\cal O}\left(\left.\frac{d^2H}{dt^2}\right|_0\right),
  \quad
  \left|
  \left.\frac{dH}{dt}\right|_0 
  \right|
  \approx
  \frac{H_0}{T_{\rm U}}
  \approx
  10^{-24}~~{\rm [1/(s~yr)]},
  \label{hubble}
\end{equation}
where $H_0 \approx 10^{-17} ~[{\rm 1/s}]$.
%Therefore $dH/dt|_0 > 0$ implies accelerating expansion of 
%the universe; otherwise, it implies decelerating expansion.
%%%%%%%%%%%%%%%%%%%%%%%%%%%%%%%%%%%%%%%%%%%%%%%%%%%%%%%%%%%%%%%%%%%%%%
\subsection{Gravitational Time Delay}
%%%%%%%%%%%%%%%%%%%%%%%%%%%%%%%%%%%%%%%%%%%%%%%%%%%%%%%%%%%%%%%%%%%%%%
The time delay functions $\Delta_e$ and $\Delta_r$ can be 
resolved into the following components:
\begin{eqnarray}
 \Delta_e (x_A, t_A, x_B) &=& 
  \hat{\Delta}(x_A, x_B) 
  + \bar{\Delta}(x_A, x_B)
  + \bar{\Delta}_e(x_A, t_A, x_B)
  + {\cal O}\left(M^2, (\dot{H}|_0)^2, \ddot{H}|_0\right),
  \label{td1}\\
 \Delta_r (x_A, x_B, t_B) &=& 
  \hat{\Delta}(x_A, x_B) 
  + \bar{\Delta}(x_A, x_B)
  + \bar{\Delta}_r(x_A, x_B, t_B)
  + {\cal O}\left(M^2, (\dot{H}|_0)^2, \ddot{H}|_0\right),
  \label{td2}
\end{eqnarray}
where $\hat{\Delta}(x_A, x_B)$ corresponds to the Shapiro time delay 
due to central mass $M$, 
$\bar{\Delta}(x_A, x_B)$ is due to the static component of cosmological 
expansion ($H_0$ in (\ref{hubble})), and
$\bar{\Delta}_e(x_A, t_A, x_B), \bar{\Delta}_r(x_A, x_B, t_B)$ 
are due to the time-dependent (secular) component of cosmological 
expansion ($dH/dt|_0$ in (\ref{hubble})).
By the straightforward calculations from (\ref{tdf1}), (\ref{tdf2}),
(\ref{rm4}), and (\ref{hubble}), we obtain following results:
\begin{eqnarray}
 \hat{\Delta}(x_A, x_B) &=&
  \frac{GM}{c^2}
  \left[
   2\ln 
   \left|\frac{x_B + \sqrt{x^2_B + b^2}}
   {x_A + \sqrt{x^2_A + b^2}}\right|
   \right.
   \nonumber\\
   & &-
    \left.
   \left(
    \frac{x_B}{\sqrt{x^2_B + b^2}} - 
    \frac{x_A}{\sqrt{x^2_A + b^2}}\right)
  \right] + {\cal O}(M^2),
    \label{td3}\\
%%%%%%%%%%%%%
 \bar{\Delta}(x_A, x_B) &=& 
 - \varepsilon\frac{H_0}{c}(x^2_B - x^2_A)
 + \frac{H_0^2}{6c^2}
  \left[
   x^3_B - x^3_A + 3b^2(x_B - x_A)
\right]
\nonumber\\
 & &- \varepsilon\frac{4GMH_0}{c^3}
  \left(\sqrt{x^2_B + b^2} - \sqrt{x^2_A + b^2}\right)
  + {\cal O}\left(M^2, (\dot{H}|_0)^2, \ddot{H}|_0\right)
  \label{td4}\\
%%%%%%%%%%%%% 
\bar{\Delta}_e(x_A, t_A, x_B) &=&
\left.\frac{dH}{dt}\right|_0
\left(- \frac{\varepsilon}{c}
\left[
(x^2_B - x^2_A)t_A + \frac{1}{3}(2x^2_B - x_A x_B - x^2_A)T_{AB}
\right]
\right.\nonumber\\
& &+ \frac{H_0}{3c^2}
 \left\{
\left[
x^3_B - x^3_A + 3b^2(x_B - x_A)
\right]t_A
\right.\nonumber\\
& &+\frac{1}{4}
\left.
 \left[
  3x^3_B - x_A x^2_B - x^2_A x_B - x^3_A + 6b^2(x_B - x_A)
 \right]T_{AB}
\right\}\nonumber\\
& &
- \varepsilon\frac{2GM}{c^3R_{AB}}
\left\{
\left[
(x_B - 2x_A)T_{AB} + 2(x_B - x_A)t_A
\right]\sqrt{x^2_B + b^2}
\right.\nonumber\\
& &
+
\left[
x_A T_{AB} - 2(x_B - x_A)t_A
\right]\sqrt{x^2_A + b^2}
\nonumber\\
& &
\left.\left.
- b^2T_{AB}\ln
\left|
\frac{x_B + \sqrt{x^2_B + b^2}}{x_A + \sqrt{x^2_A + b^2}}
\right|
\right\}
\right)
+ {\cal O}\left(M^2, (\dot{H}|_0)^2, \ddot{H}|_0\right)
\label{td5}\\
%%%%%%%%%%%%%%%
 \bar{\Delta}_r(x_A, x_B, t_B) &=&
\left.\frac{dH}{dt}\right|_0
\left(- \frac{\varepsilon}{c}
\left[
(x^2_B - x^2_A)t_B - \frac{1}{3}(x^2_B + x_A x_B - 2x^2_A)T_{AB}
\right]
\right.\nonumber\\
& &+ \frac{H_0}{3c^2}
 \left\{
\left[
x^3_B - x^3_A + 3b^2(x_B - x_A)
\right]t_B
\right.\nonumber\\
& &-
\frac{1}{4}
\left.
 \left[
  x^3_B + x_A x^2_B + x^2_A x_B - 3x^3_A + 6b^2(x_B - x_A)
 \right]T_{AB}
\right\}\nonumber\\
& &
+ \varepsilon\frac{2GM}{c^3R_{AB}}
\left\{
\left[
x_BT_{AB} - 2(x_B - x_A)t_B
\right]\sqrt{x^2_B + b^2}
\right.\nonumber\\
& &
-
\left[
(2x_B - x_A) T_{AB} - 2(x_B - x_A)t_B
\right]\sqrt{x^2_A + b^2}
\nonumber\\
& &
\left.\left.
- b^2T_{AB}\ln
\left|
\frac{x_B + \sqrt{x^2_B + b^2}}{x_A + \sqrt{x^2_A + b^2}}
\right|
\right\}
\right)\nonumber\\
& &
+ {\cal O}\left(M^2, (\dot{H}|_0)^2, \ddot{H}|_0\right),
\label{td6}
\end{eqnarray}
where $\varepsilon = N^x_{AB} = 1$ for $x_B - x_A > 0$ and
$\varepsilon = - 1$ for $x_B - x_A < 0$. $\varepsilon$ is derived 
from the term $ - 2N^i_{AB}g^{0i}_{(1)}$ in (\ref{tdf1}) and (\ref{tdf2}).
%%%%%%%%%%%%%%%%%%%%%%%%%%%%%%%%%%%%%%%%%%%%%%%%%%%%%%%%%%%%%%%%%%%%%%%
\subsection{Equation of Light Time}
%%%%%%%%%%%%%%%%%%%%%%%%%%%%%%%%%%%%%%%%%%%%%%%%%%%%%%%%%%%%%%%%%%%%%%
Actually, $x_A$ and $x_B$ are positions at
$t = t_A$ and $t = t_B$, respectively:
\begin{eqnarray}
 x_A = x_A(t_A),\quad
  x_B = x_B(t_B).
\end{eqnarray}
Then, (\ref{td1}) and (\ref{td2}) can be symbolically expressed as
\begin{eqnarray}
 \Delta_e (x_A(t_A), t_A, x_B(t_B)) &=& 
  \hat{\Delta}(x_A(t_A), x_B(t_B))
  + \bar{\Delta}(x_A(t_A), x_B(t_B))
  \nonumber\\
  & &+ \bar{\Delta}_e(x_A(t_A), t_A, x_B(t_B)),
  \label{elt1}\\
 \Delta_r (x_A(t_A), x_B(t_B), t_B) &=& 
  \hat{\Delta}(x_A(t_A), x_B(t_B))
  + \bar{\Delta}(x_A(t_A), x_B(t_B))
  \nonumber\\
  & &+ \bar{\Delta}_r(x_A(t_A), x_B(t_B), t_B),
  \label{elt2}
\end{eqnarray}
Hence, the equations
\begin{eqnarray}
 t_B - t_A &=& 
  \frac{1}{c}
  \left[
   R_{AB}(x_A(t_A), x_B(t_B)) 
  + \Delta_e (x_A(t_A), t_A, x_B(t_B))
  \right],
  \label{elt3}\\
 t_B - t_A &=& 
  \frac{1}{c}
  \left[
   R_{AB}(x_A(t_A), x_B(t_B)) 
  + \Delta_r (x_A(t_A), x_B(t_B), t_B)
  \right],
  \label{elt4}
\end{eqnarray}
can be considered as the equation of light time.
Here that $x_A$ and $x_B$ must be known beforehand as a function of 
time $t$ via the numerical integration of the equation of motion 
or via lunar-planetary ephemerides.
\subsection{Time-dependency of Solution}
\begin{figure}
   \begin{center}
    \includegraphics[scale=0.3]{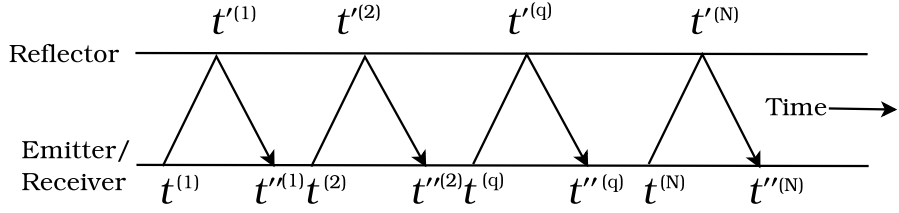}
   \end{center}
 \caption{Time-dependency of solutions. The emission time,
 reflection time, and reception time 
 of the $(q)$-th observation are denoted by
 $t^{(q)}$, $t^{\prime (q)}$, and $t^{\prime\prime (q)}$, 
 respectively. \label{fig3}}
\end{figure}

%%%%%%%%%%%%%%%%%%%%%%%%%%%%%%%%%%%%%%%%%%%%%%%%%%%%%%%%%%%%%%%%%%%%
The solutions from (\ref{td1}) to (\ref{td6}) have the 
following form:
\begin{eqnarray}
  {\cal T}^{(q)}_e(x^{(q)}_A, t^{(q)}_A, x^{(q)}_B)
  &=& 
  \frac{1}{c}
  \left[
   R^{(q)}_{AB}(x^{(q)}_A, x^{(q)}_B)
  + E_1^{(q)}(x^{(q)}_A, x^{(q)}_B)
  \right.
  \nonumber\\
 & &
  \left.
  + E_2^{(q)}(x^{(q)}_A, x^{(q)}_B)t^{(q)}_A
  \right],
  \label{td7}\\
  {\cal T}^{(q)}_r(x^{(q)}_A, x^{(q)}_B, t^{(q)}_B)
  &=& 
  \frac{1}{c}
  \left[
  R^{(q)}_{AB}(x^{(q)}_A, x^{(q)}_B)
  + R_1^{(q)}(x^{(q)}_A, x^{(q)}_B)
  \right.
  \nonumber\\
 & &
  \left.
   + R_2^{(q)}(x^{(q)}_A, x^{(q)}_B)t^{(q)}_B
  \right],
  \label{td8}
\end{eqnarray} 
where the superscript $(q)$ implies the $q$-th observation,
$E_1^{(q)}(x^{(q)}_A, x^{(q)}_B), R_1^{(q)}(x^{(q)}_A, x^{(q)}_B)$
constitute the static component of gravitational time delay, and
$E_2^{(q)}(x^{(q)}_A, x^{(q)}_B)t_A, 
R_2^{(q)}(x^{(q)}_A, x^{(q)}_B)t_B$ 
constitute the component depending on either $t_A$ or $t_B$.
If we carry out a series of observations, $q = 1, 2, 3, \cdots, N$
(see Fig \ref{fig3}), $t^{(q)}_A, t^{(q)}_B$ on the right-hand side 
of (\ref{td7}) and (\ref{td8}) can be regarded as the time-dependent 
(secular) component of the solution. 
%%%%%%%%%%%%%%%%%%%%%%%%%%%%%%%%%%%%%%%%%%%%%%%%%%%%%%%%%%%%%%%%%%%%%%%%%
\section{Application to Secular Increase in Astronomical Unit
\label{siau}}
%%%%%%%%%%%%%%%%%%%%%%%%%%%%%%%%%%%%%%%%%%%%%%%%%%%%%%%%%%%%%%%%%%%%%%%%%
The Secular increase in the astronomical unit (AU), reported by 
Krasinsky and Brumberg (2004) \cite{kb2004}, is an unexplained 
physical phenomenon observed in the solar system. This anomaly was 
discovered while analyzing planetary radar and spacecraft 
(mainly Martian landers/orbiters) ranging data and improving the 
various astronomical constants including AU. 
\cite{kb2004} estimated
\begin{equation}
\frac{d{\rm AU}}{dt} = 15 \pm 4 ~{\rm [m/cy]}.
\end{equation}
as the most appropriate value. 
Subsequently $d{\rm AU}/dt \simeq 20 ~{\rm [m/cy]}$ was
separately evaluated by Pitjeva at the Institute of Applied Astronomy
(IAA), Russia, and by Standish at the Jet Propulsion Laboratory
(JPL), USA 
\footnote{\cite{kb2004} used only the radiometric 
observational data, whereas Pitjeva and Standish carried out  
analysis using various kinds of data such as  optical and VLBI; 
each of them worked with different software. However, as it 
stands, their results have not been published officially because of 
the difficulty in interpreting the obtained value, i.e., 
$20 ~{\rm [m/cy]}$ \cite{pitjeva2009}.}.
%
%It should be emphasized that the increase in AU does not imply
%the expansion of a planetary orbit or the equivalent increase in 
%the orbital period of a planet. According to Krasinsky \cite{krasinsky}, 
%the observations do not reveal any such traces. Further, 
%the determination error of inner planetary orbits in the 
%latest lunar-planetary ephemerides (e.g. DE of JPL and EPM of IAA) 
%is also smaller than the observed $d{\rm AU}/dt$, i.e., 
%15 [m/cy] (see Table 4 of \cite{pitjeva2005}). Therefore, the 
%observed $d{\rm AU}/dt$ value may relate with not the 
%dynamic aspect of planetary motion but the 
%propagation of a light ray/signal.

Previously, some attempts have been made to explain the 
secular trend in terms of cosmological expansion 
\cite{kb2004,mashhoon,arakida2010}.
In particular, \cite{kb2004,arakida2009} 
considered its contribution to light propagation.
However, as mentioned in Section \ref{intro}, it is generally
difficult to compute the time-dependent geodesic of null rays; 
hence, the approach of the former is somewhat 
qualitative, whereas that of the latter is essentially 
restricted to discussion in static spacetime. 

Now, on the basis of the results obtained in previous section, 
let us re-examine whether the cosmological effect relates with the observed 
$d{\rm AU}/{dt}$. It is appropriate to regard the coefficients of 
$t_A$ and $t_B$ in (\ref{td7}) and (\ref{td8}) as secular 
terms owing to cosmological expansion. 
Then, if we assume that 
$dH/dt|_0 \approx 10^{-24} ~[{\rm 1/(s~ yr)}]$ from (\ref{hubble}),
the leading order of magnitude of coefficients is,
\begin{eqnarray}
 E_2(x_A, x_B) \approx R_2(x_A, x_B) \approx 
  10^{-10} ~~{\rm [m/yr]} = 10^{-8} ~~{\rm [m/cy]}
\end{eqnarray}
because $x_A$ and $x_B$ are of the order of a few [AU] or 
$10^{11} ~[{\rm m}]$ in the solar system.
Unfortunately, this is approximately 9 orders of magnitude 
smaller than the evaluated $d{\rm AU}/dt$, i.e., 
$ 15 ~[{\rm m/cy}]$.
Therefore, the time-dependent effect due to cosmological
expansion does not induce the secular increase in AU.
%%%%%%%%%%%%%%%%%%%%%%%%%%%%%%%%%%%%%%%%%%%%%%%%%%%%%%%%%%%%%%%%%%%%%%
\section{Summary\label{concl}}
%%%%%%%%%%%%%%%%%%%%%%%%%%%%%%%%%%%%%%%%%%%%%%%%%%%%%%%%%%%%%%%%%%%%%%
We considered light propagation in the time-dependent 
McVittie spacetime, which can be considered as the 
spacetime around a gravitating body such as the Sun, embedded in the 
FLRW cosmological background metric. To discuss the time-dependent
null ray, we adopted the recently developed time transfer function 
method \cite{plt2004,pt2008} which is originally related to 
Synge's world function $\Omega(x_A, x_B)$ and precludes the
integration of the null geodesic equation. The first step 
involved the application of this method to McVittie spacetime and 
re-examination of the cosmological effect on the round-trip time of 
a light ray/signal in the solar system. 
The time delay functions $\Delta_e$ and $\Delta_r$, which characterize 
the gravitational time delay, were given by the functions of not only
the spacial coordinates $x_A $ and $x_B$ but also the emission 
time $t_A$ or reception time $t_B$ (see (\ref{td1} to (\ref{td6}));
the presence of the terms $t_A$ and $t_B$ in $\Delta_e$ and $\Delta_r$ 
express the time-dependency of the solution.

On the basis of the obtained results, we also investigated the 
secular increase in the astronomical unit (AU), reported by 
\cite{kb2004}; we showed that the 
leading order terms of the time-dependent component due to cosmological 
expansion are 9 orders of magnitude smaller than the observed 
value of $d{\rm AU}/dt$, i.e., $ 15 \pm 4$ ~[m/cy].
Therefore, we explicitly asserted that it is not possible to explain 
the secular increase of AU in terms of cosmological expansion.

Currently, it is assumed that the most plausible cause of 
$d{\rm AU}/dt$ is either the lack of calibrations of 
internal delays in the radio signals within spacecraft or 
the complexity in modeling the solar corona; however, 
no conclusive evidence has not been reported thus far. The origin of 
the secular increase in AU has also been examined in
terms of other physical aspects such as solar mass loss 
\cite{kb2004,noerdlinger}, the time variation of the gravitational 
constant $G$ \cite{kb2004}, the influence of dark matter in the 
solar system \cite{arakida2010}, the multi-dimensional brane world 
scenario \cite{iorio2005}, the transfer of rotational angular momentum 
of the Sun due to solar mass loss \cite{miura2009}, the 
azimuthally symmetric theory of gravitation \cite{nyambuya}, and 
the kinematics in Randers-Finsler geometry \cite{lichang}. 
%Some of them provided interesting explanations in terms of 
%planetary motion, i.e., the mechanism of expansion of planetary 
%orbits. However, thus far, there is hardly any trace of the 
%expansion of planetary orbits because the determination error of 
%inner planetary orbits in the latest ephemerides is 
%approximately 1 [m] or less (see Table 4 of \cite{pitjeva2005}). 
However, so far the cause of the secular increase in AU is not clear.
Nonetheless, this phenomenon and other anomalies 
discovered in the solar system may be related to some 
fundamental properties of gravitation; therefore it is important to 
verify these phenomena theoretically and observationally.

On the other hand, the cosmological effect on gravitationally bound 
local systems has been widely investigated, see i.e. 
\cite{carrera} and references therein. 
Its influence in the solar system is probably so 
small that one cannot detect it presently. 
Nevertheless, it seems that the theoretical discussion is not 
resolved. Therefore, from the theoretical point of view, 
it is interesting to construct robust theoretical model beyond 
the McVittie model and bring an end to the argument consistently. 

Owing to the rapid enhancement of observational techniques,
more accurate and rigorous formulae are required to express
light propagation. \cite{plt2004} and \cite{pt2008} succeeded in 
developing an elegant and useful method for constructing the 
post-Newtonian/post-Minkowskian approximation, and they showed 
that Synge's world function $\Omega(x_A, x_B)$ (which is not widely 
adopted or discussed) is a powerful tool for formulations in practical 
problems. Thus, the value of Synge's world function can be
reaffirmed, and it is possible to proceed with the theoretical 
development of observational models based on it. 

\begin{acknowledgements}
We acknowledge the anonymous referees for fruitful comments and
suggestions. We would like to extend his gratitude to Prof. 
G. A. Krasinsky, Dr. E. Yagudina, and Dr. E. V. Pitjeva for their 
valuable discussions, suggestions, and hospitality during author's 
visit to the Institute of Applied Astronomy in St. Petersburg, Russia. 
We also thank Dr. C. Le Poncin-Lafitte; his work strongly influenced 
the present study. This work was partially supported by the Ministry of 
Education, Culture, Sports, Science and Technology (MEXT), Japan, via
Grant-in-Aid No. 21740193.
\end{acknowledgements}

% BibTeX users please use one of
%\bibliographystyle{spbasic}      % basic style, author-year citations
%\bibliographystyle{spmpsci}      % mathematics and physical sciences
%\bibliographystyle{spphys}       % APS-like style for physics
%\bibliography{}   % name your BibTeX data base

% Non-BibTeX users please use

\end{document}